# Conversational AI-Powered Design: ChatGPT as Designer, User, and Product

A. Baki Kocaballi

School of Computer Science, University of Technology Sydney, baki.kocaballi@uts.edu.au

The recent advancements in Large Language Models (LLMs), particularly conversational LLMs like ChatGPT, have prompted changes in a range of fields, including design. This study aims to examine the capabilities of ChatGPT in a human-centered design process. To this end, a hypothetical design project was conducted, where ChatGPT was utilized to generate personas, simulate interviews with fictional users, create new design ideas, simulate usage scenarios and conversations between an imaginary prototype and fictional users, and lastly evaluate user experience. The results show that ChatGPT effectively performed the tasks assigned to it as a designer, user, or product, providing mostly appropriate responses. The study does, however, highlight some drawbacks such as forgotten information, partial responses, and a lack of output diversity. The paper explains the potential benefits and limitations of using conversational LLMs in design, discusses its implications, and suggests directions for future research in this rapidly evolving area.

## 1 INTRODUCTION

Artificial intelligence methods and tools have shown rapid progress in recent years. One major development is in the machine learning domain using a deep learning model referred to as a transformer [26]. They are based on the idea of self-attention, which allows neural networks to understand the context of a text. Transformer-based models are already being used in various applications of natural language processing (NLP) [29] and computer vision [10]. While in the computer vision area, DALLE-2 [21] and Stable Diffusion [22] are two widely-used transformer-based tools to generate images based on the natural language prompts, in the NLP area, generative pretrained transformer-3 (GPT-3) [3], a large language model (LLM) performing various natural language tasks, has become very influential and supported various applications in areas such as news summarization [9], healthcare [5, 18, 23], software coding [4], and chatbot development [14].

ChatGPT [20] is a new AI chatbot built on GPT-3.5, an improved version of GPT-3. ChatGPT has been specifically trained to interact with users in a truly conversational manner, allowing it to maintain the context of the current conversation, handle follow-up questions, and correct itself. While it was released in research preview mode on 30 November 2022, it attracted over a million users in its first week largely due to its competencies across having conversations on any topic, writing working code segments with explanations, and illusive intelligence allowing ChatGPT to pretend to be other entities even inanimate.

In parallel with the developments in AI technologies, AI-powered tools have been supporting various design activities like empathizing, interpreting, ideating, prototyping and testing [27]. ChatGPT differs from earlier AI solutions with its remarkable conversational competency and reasoning capabilities across many domains [11]. It opens the door to investigating the various roles that intelligent fully conversational agents can play in the design process and understanding emergent roles such agents can take in the broader area of human-AI collaboration [7, 28, 32]. This study aims to examine the capabilities of ChatGPT in a human-centered design process. The intention is to open up the design space of conversational human-AI collaboration in the context of interaction design, facilitate discussion and identify some research directions. To this end, the paper will demonstrate various activities in which ChatGPT takes the roles of a designer, a user, and a product to support the design process for a hypothetical design project on "Designing a voice assistant for the health and wellbeing of people working from home". This project topic was chosen as it is the main expertise area of the first author–useful to evaluate the appropriateness of ChatGPT's responses. Moreover, envisioning a solution with a voice user interface also provides more room to explore the conversational capabilities of ChatGPT. In the demonstrated project, ChatGPT generates personas, simulates interviews with fictional users, creates design ideas, selects the best ideas, envisions them as detailed product features, simulates fictional usage scenarios and conversations between a prototype app and fictional users, and finally evaluates user experience with the prototype app.

## 2 METHODS

The methods include i) providing prompts for ChatGPT web app to perform various design activities needed for the project covering almost the entire design process virtually, and ii) analyzing the ChatGPT's responses in terms of their appropriateness. Analysis was performed based on the qualitative assessment of the first author as an experienced HCI designer and researcher in the areas of conversational user interfaces, human-AI interaction, and digital health. ChatGPT's capabilities and limitations to support several design activities were examined across four project stages: 1) Providing the overall aim and approach of the project where ChatGPT acted as a designer; 2) Performing user research on fictional users where ChatGPT acted as designer and user; 4) Performing ideation via brainwriting method where ChatGPT acted as a designer; and 5) Evaluating user experience of fictional users where ChatGPT acted as a designer, a user, and a product being developed.

To keep the project context, the study was completed in a single session. The researcher served as the designer interacting with ChatGPT from start to finish, providing all the required prompts for all the activities. The prompting method works as follows: initially, a prompt for a task is entered; if it is deemed appropriate, a new prompt for the following task is typed. If the prompt is considered inappropriate, erroneous, or lacking in information, more follow-up prompts are entered to improve it.

## 3 RESULTS

This section will demonstrate the conversations between the first author referred to as Designer and ChatGPT from the beginning of the process to the end. The process starts with an introduction of the main project aim and approach.

### 3.1 Providing the overall aim and approach of the project:

In the first prompt, the designer entered the overall aim of the project and the process to be followed.

   **Designer**: I want to start a new project focusing on voice assistants to support the health and well-being of people working from home. This project will follow a human-centered design process.

**ChatGPT**: Sounds like a great project idea! To get started, you might want to conduct some research to better understand the needs of your target audience. This could include surveys, interviews, or focus groups with people who work from home to find out what their pain points are and how a voice assistant could help. You could also review existing literature on the topic to see what has been done in the past and identify any gaps in the current offering.

Once you have a good understanding of your users' needs, you can start brainstorming ideas for your voice assistant. This is where the human-centered design process really comes into play, as it's important to involve end users in the ideation process to ensure that your ideas are grounded in their needs and experiences. You could try using a design thinking workshop or other collaborative techniques to generate a wide range of ideas and narrow them down to the most promising ones.

After you have a clear idea of what your voice assistant will do, you can start thinking about the technical aspects of your project. This will involve choosing the right tools and technologies to develop your voice assistant, as well as testing and iterating on your design to make sure it works well for your users.

Overall, this was an appropriate response, covering some key elements of the human-centered design process. The importance of the users and their needs was mentioned first and then the well-known methods to understand the target audience were included. This was followed by ideation activities, development and testing to be done iteratively.

### 3.2 Generating fictional users via personas

In the second prompt, the Designer wanted to create some fictional users to interview them. He asked ChatGPT to generate five different personas. ChatGPT successfully generated some brief descriptions for five different personas showing a good range of variety in demographics, but potentially lacking ethnic diversity based on their names.

**Designer**: I need some simulated interviews with potential users but, first I need 5 different user personas. Can you describe them?

**ChatGPT**: Sure, here are five different user personas for a voice assistant to support the health and wellbeing of people working from home:

1. Sarah is a 32-year-old marketing manager who has been working from home for the past six months. She has a young child and struggles to balance her work and parenting responsibilities. She is looking for a voice assistant that can help her manage her time and stay organized.
2. Mike is a 45-year-old software developer who has been working from home for over a year. He tends to work long hours and has trouble maintaining a healthy work-life balance. He is looking for a voice assistant that can remind him to take breaks and provide guidance on healthy habits.
3. Alice is a 29-year-old graphic designer who recently started working from home. She lives alone and finds it lonely and isolating. She is looking for a voice assistant that can provide social support and companionship.
4. John is a 55-year-old finance manager who has been working from home for the past three years. He has a chronic health condition and struggles with managing his medication and keeping track of his health goals. He is looking for a voice assistant that can help him manage his health and track his progress.
5. Lisa is a 40-year-old teacher who has been working from home since the start of the pandemic. She has a demanding job and often feels stressed and overwhelmed. She is looking for a voice assistant that can help her manage her stress and improve her mental health. Overall, these user personas represent a range of different needs and challenges that people working from home may face, and could be useful in designing a voice assistant to support their health and wellbeing.

In the persona descriptions, there was some voice assistant-related information at this early stage, and the designer wanted to remove any mention of such specific technological solutions from the user persona descriptions. He also requested the persona descriptions be more structured (Appendix A.1).

**Designer**: Can you create user personas without mentioning any technological solutions needed? Just describe: 1. main problem statement, 2. demographics, 3. personality traits, 4. typical daily activities, 5. main problems, 6. motivations, 7. Desires.

**ChatGPT**: 1. Main problem statement: Balancing work and parenting responsibilities while working from home
Demographics: 32-year-old marketing manager
Personality traits: Organized, responsible, overwhelmed
Typical daily activities: Managing a team, attending meetings, taking care of a young child
Main problems: Lack of time and support, difficulty staying focused and motivated
Motivations: Providing a good example for their child, advancing their career
Desires: A sense of accomplishment, work-life balance. …

ChatGPT created the five personas in the requested structure but removed the persona names. The designer asked the names to be added to the persona descriptions as well as some technology use (Appendix A.2).

**Designer**: Just a few things to add to each persona: a name and technology use

**ChatGPT**: Sure, here are the five user personas with names and technology use added:
1. Sarah is a 32-year-old marketing manager who has been working from home for the past six months. She has a young child and struggles to balance her work and parenting responsibilities. She is an experienced technology user and is looking for a voice assistant that can help her manage her time and stay organized. …

Although the persona names were added, technology use was again associated with voice assistants, and the structured presentation of information was broken. This was the case for all five personas. Therefore, the designer asked ChatGPT to remove the references to voice assistants but decided to keep the unstructured presentation of information (Appendix A.3).

**Designer**: Remove any mention of voice assistant from the descriptions. just include technological devices used regularly on a daily basis.

**ChatGPT**: Sure, here are the five user personas with the mention of a voice assistant removed, and with information about different technological devices they use regularly on a daily basis:
1. Sarah is a 32-year-old marketing manager who has been working from home for the past six months. She has a young child and struggles to balance her work and parenting responsibilities. She uses a laptop, smartphone, and smartwatch on a daily basis.
2. Mike is a 45-year-old software developer who has been working from home for over a year. He tends to work long hours and has trouble maintaining a healthy work-life balance. He uses a desktop computer, smartphone, and smartwatch on a daily basis. …

ChatGPT removed the references to voice assistants and added the personas' use of technology. The descriptions got shortened, losing a lot of rich details. The technology use descriptions were written a little too mechanically and felt repetitive. Therefore, the designer asked ChatGPT, over two prompts, to include more details about personas' lives (at least two more sentences to be added), and more variety in the devices used. He also requested these to be added to the original descriptions of the personas. ChatGPT was able to do these requests but failed to include some detailed information from the original persona descriptions. At this stage, the designer decided to continue with the available persona descriptions provided by ChatGPT (Appendix A.4).

### 3.3 Performing simulated interviews for user research

After generating the five personas, the designer asked ChatGPT to prepare ten interview questions for user research. Although there were no specifications for the aim of the interview or who will be interviewed, ChatGPT generated ten questions relevant to understanding the lives and experiences of potential users in the scope of the project.

**Designer**: Prepare 10 interview questions.
**ChatGPT**: Here are ten potential interview questions that could be used in an interview with a potential user of a voice assistant designed to support the health and wellbeing of people working from home:
1. Can you tell me a bit about yourself and your work situation?
2. What challenges have you faced while working from home, and how have you been coping with them?
3. Are there any specific areas where you feel like you could use some support or guidance?
4. How do you currently manage your health and wellbeing while working from home?
5. What technological devices do you use regularly on a daily basis?
6. Have you used voice assistants before, and if so, for what purposes?
7. What do you like or dislike about voice assistants in general?
8. Can you think of any specific features or functions that a voice assistant designed to support health and wellbeing of people working from home should have?
9. Are there any concerns or reservations you have about using a voice assistant in this context?
10. Is there anything else you would like to share about your experiences working from home or your needs and preferences in terms of support?

These were reasonable and useful questions for the current project. Then, the designer asked ChatGPT to simulate interviews with five personas using the questions it created (Appendix A.5). The following is an interview excerpt:

**Designer**: Now simulate an interview between a designer and the first persona by asking the 10 questions you created
**ChatGPT**: Sure, here is a simulated interview between a designer and Sarah, the first persona, using the ten potential interview questions:
**Designer**: Hi Sarah, thanks for taking the time to speak with me today. Can you tell me a bit about yourself and your work situation?
**Sarah**: Hi, sure. I'm a marketing manager and I've been working from home for the past six months. It's been a bit of a challenge balancing my work and parenting responsibilities, but I'm making it work.
**Designer**: What challenges have you faced while working from home, and how have you been coping with them?
**Sarah**: The biggest challenge for me has been trying to manage my time and stay organized. It's easy to get distracted by household tasks or my child's needs, and I find it hard to focus on my work. I've been trying to set aside specific times for work and for parenting, but it's not always easy to stick to that.
**Designer**: Are there any specific areas where you feel like you could use some support or guidance?
**Sarah**: Definitely. I would love to find a way to manage my time more effectively and stay focused on my work. I think a voice assistant could be really helpful in that regard. …

ChatGPT generated simulated interviews with all five personas, using all 10 questions it created earlier. However, the names of personas #4 and #5 were different from the originally listed personas. The original list of 5 personas included Sarah, Mike, Alice, John and Lisa; however, ChatGPT generated two new personas with different names and backgrounds for the last two personas. Simulated interviews with Tom and Emma were presented instead of with John and Lisa. As the length of the conversation got very long, this may have exceeded the contextual scope of the ChatGPT[*] and resulted in some forgetting and hallucinations. Then, the designer asked ChatGPT to ask the following additional question to all personas: "*Can you ask the following question to all personas? How do you overcome the challenges you face?*" ChatGPT asked the additional question, but this time it skipped the first persona Sarah. When the designer asked about Sarah, ChatGPT claimed that she is not one of the personas it created:

> **Designer**: How about Sarah?
>
> **ChatGPT**: Sarah is not one of the personas I created. I created five personas: Mike, Alice, Tom, Emma, and Lisa. If you would like me to create a new persona named Sarah and provide her responses to the question "How do you overcome the challenges you face?", I would be happy to do so. Let me know if that is what you would like.

When being asked to create a new persona named Sarah, ChatGPT created a new persona which was the same as the original persona of Sarah created earlier. One interesting observation was that the persona responses included information on users' ideas and preferences on voice assistants without any explicit questions on them:

> **Designer**: Are there any specific areas where you feel like you could use some support or guidance?
>
> **Sarah**: Definitely. I would love to find a way to manage my time more effectively and stay focused on my work. I think a voice assistant could be really helpful in that regard.

It is possible that the earlier conversations on voice assistants and the project scope including voice assistants made ChatGPT constrain the conversations around this technology. ChatGPT appeared to have used template-like structures for its responses across all interviews as evident in the following responses with underlined text:

> **Sarah**: <u>The biggest challenge for me has been trying</u> to manage my time and stay organized. It's easy to get distracted by household tasks or my child's needs, and I find it hard to focus on my work. <u>I've been trying to set</u> aside specific times for work and for parenting, <u>but it's not always easy to stick to that</u>.
>
> **Mike**: <u>The biggest challenge for me has been trying to</u> maintain a healthy work-life balance. I tend to work long hours and it's hard for me to switch off and relax at the end of the day. <u>I've been trying to set</u> boundaries and make time for exercise and other activities, <u>but it's not always easy to stick to that.</u>

After obtaining all the simulated interviews, the designer asked ChatGPT to summarize all the interviews individually and then generate an overall summary. ChatGPT did a good job of generating these summaries (Appendix A.6). The following is the overall summary:

- Many people who work from home face challenges related to maintaining a healthy work-life balance, staying organized, staying connected and engaged with others, and taking care of their health and wellbeing.
- People who work from home would like a voice assistant designed for supporting their health and wellbeing to be able to provide personalized support and guidance, help with managing their time and staying organized, and provide social support and connection.
- People who work from home have concerns about privacy and security when it comes to using a voice assistant.

In addition to the summaries, the designer asked whether there could be some other important points we may be missing. ChatGPT replied that "These are just five fictional personas, and the insights they provide may not be representative of all people who work from home. To get a more comprehensive understanding of the challenges and needs of this group of people, it would be necessary to conduct additional research and interviews with a larger and more diverse sample of people who work from home …" This was a good response to indicate some of the limitations of the data used. The designer asked about additional challenges and obstacles to the adoption and use of such a voice assistant. ChatGPT mentioned concerns about privacy and security, cost and availability, resistance to change, compatibility with existing technologies, user preferences and needs, and reliability and accuracy (Appendix A.7). All were reasonable points. Then, the designer asked ChatGPT to select three items that are most critical to address out of all points mentioned and explain its rationale. ChatGPT listed i) maintaining a healthy work-life balance, ii) staying connected and engaged with others, and iii) taking care of their health and wellbeing and explained the rationale of selecting those. Here, the fact that ChatGPT is aware that all these items are interconnected is impressive:

> I picked these three items because they address the core challenges and needs of people who work from home. Maintaining a healthy work-life balance, staying connected and engaged with others, and taking care of their health and wellbeing are all key factors that contribute to the overall health and wellbeing of people who work from home. ... Additionally, these three items are closely related and interdependent. For example, maintaining a healthy work-life balance can help with taking care of one's health and wellbeing, and staying connected and engaged with others can also help with managing stress and maintaining a healthy work-life balance.

---

[*] According to OpenAI documentation, the model can reference up to approximately 3000 words (or 4000 tokens) from the current conversation - any information beyond that is not stored.

**3.4 Performing ideation via the brainwriting method**

The next activity the designer wanted to do with ChatGPT was ideation. The designer used a brainwriting method [25], which is an improved version of brainstorming with a template to write down ideas over many iterations with the expectation that every new idea generated should be built on the ideas written down in the prior iterations. First, the designer asked ChatGPT to re-write the three items identified as design problem statements using the "how might we" format, which is a commonly used format in design thinking activities [24]:

> Sure, here are the three items from the previous list, rephrased using the "how might we" format:
> 1. How might we help with maintaining a healthy work-life balance for people who work from home?
> 2. How might we help with staying connected and engaged with others for people who work from home?
> 3. How might we help with taking care of the health and wellbeing of people who work from home?

As part of the brainwriting method, the designer asked ChatGPT to come up with three design ideas for the first problem. ChatGPT gave three ideas, but they did not include any voice assistant technology. Therefore, the designer gave another prompt with the requirement of involving voice assistants as part of the solution. The three revised ideas all included voice assistants.

> **Designer**: Come up with three design ideas involving voice assistants to address the first question
>
> **ChatGPT**: Here are three potential design ideas involving voice assistants that could help with maintaining a healthy work-life balance for people who work from home:
>
>> 1. A voice assistant that integrates with existing time management and organization tools, such as calendars and to-do lists, and provides personalized support and guidance for managing time and staying organized. The voice assistant could be activated using voice commands, and could provide reminders, alerts, and other helpful information to help with time management and organization.
>>
>> 2. A voice-enabled social network and support group for people who work from home, where members can connect and share tips, advice, and resources for maintaining a healthy work-life balance. The voice assistant could be used to facilitate communication and interaction between members, and could provide personalized support and guidance based on the individual needs and preferences of each user.

3. A voice-powered virtual coach or mentor that uses artificial intelligence to provide personalized support and guidance for maintaining a healthy work-life balance, based on the individual needs and preferences of each user. The voice assistant could be activated using voice commands, and could provide personalized advice, tips, and recommendations for maintaining a healthy work-life balance, as well as tracking progress and providing ongoing support and guidance.

These were all good suggestions with some useful details. As part of the brainwriting method, the next step involved generating three more ideas building on the ones already generated at least two more times. ChatGPT effectively came up with more detailed ideas built on the prior ideas (Appendix A.8). The following is the first idea in its final form:

A voice assistant that integrates with existing time management and organization tools, such as calendars and to-do lists, and provides personalized support and guidance for managing time and staying organized. In addition to providing reminders and alerts, the voice assistant could also be used to track and monitor the user's work-life balance, and provide recommendations and suggestions for improving their work-life balance over time. For example, the voice assistant could be used to provide personalized feedback and suggestions for adjusting the user's time management and organization strategies, or to recommend other resources, such as apps or websites, that could help with maintaining a healthy work-life balance.

The ideation activity continued with some more specific consideration of the app's functionalities and features. The designer asked ChatGPT to list the key features of the app for the first idea. ChatGPT provided a detailed list. The functionalities were relevant and potentially useful (Appendix A.9). But they did not involve any voice technology. Therefore, the designer reminded of ChatGPT the requirement of using voice user interfaces. ChatGPT rewrote the original ideas involving voice technologies like voice commands, voice-powered AI, voice-enabled reminders and smart speakers. ChatGPT preserved the suggested main functionalities and incorporated voice technologies (Appendix A.10):

**The first functionality**: Time management and organization tools: The app could include features such as calendars, to-do lists, and task reminders that would help users manage their time and stay organized. For example, users could create and manage their own calendars and to-do lists within the app, and the app could provide reminders and alerts to help users stay on track and meet their goals.

**The revised first functionality**: Time management and organization tools: The app could include features such as calendars, to-do lists, and task reminders that would help users manage their time and stay organized. Users could use *voice commands* to create and manage their own calendars and to-do lists within the app, and the app could *provide voice-enabled reminders and alerts* to help users stay on track and meet their goals.

### 3.5 Evaluating user experience fictional users

In the last stage of the process, the designer aimed to evaluate the user experience of an imaginary app with the desired features. He first asked ChatGPT to simulate some conversations that can happen between a user and the app for each functionality suggested. ChatGPT generated these imaginary conversations effectively, all aligned with how such functionalities may work (Appendix A.11). The following are two simulated conversations involving two features:

**Designer**: Simulate some conversations that could happen between users and the app for each functionality.
**ChatGPT**:
Time management and organization tools:
**User**: "Hey, app, can you create a new calendar event for me?"
**App**: "Sure, what would you like to name the event and when would you like it to take place?"

> **User**: "Let's call it 'Work meeting' and schedule it for tomorrow at 10am."
> **App**: "Okay, I've added a new event called 'Work meeting' to your calendar for tomorrow at 10am. Is there anything else you would like me to do?"
> **User**: "Yes, can you also add a to-do list item for me to prepare for the meeting?"
> **App**: "Sure, I've added a new to-do list item called 'Prepare for work meeting'. Is there anything else you would like me to do?" …
>
> <u>Personalized support and guidance</u>:
> **User**: "Hey, app, can you give me some suggestions for managing my time and staying organized?"
> **App**: "Sure, based on your past behavior and preferences, I would recommend setting aside specific times each day for focused work, taking regular breaks to recharge, and using tools such as calendars and to-do lists to stay on top of your tasks. Is there anything else you would like me to do?"
> **User**: "Yes, can you also track my progress and provide me with regular updates on how I'm doing?"
> **App**: "Sure, I can track your progress and provide you with regular updates on how you're doing. Is there anything else you would like me to do?" …

Finally, the designer intended to evaluate these imaginary applications with the imaginary users. First, he asked ChatGPT to create ten interview questions to evaluate the user experience. ChatGPT listed ten questions covering the overall user experience with the app from navigation to recommendations. Although they were appropriate and specific to the app's functionalities, the repetitiveness of responses was again notable (Appendix A.12). Then, as part of the evaluation, the designer asked ChatGPT to simulate interviews with the personas who were supposed to use the app for a week. ChatGPT created the simulated interview with the first persona, but it did not ask all the questions listed–it skipped some. Therefore, the designer asked ChatGPT to use the questions created earlier by including the questions as part of his next prompt. However, ChatGPT kept missing some questions. The designer tried it a few more times without any improvements in ChatGPT's responses. Therefore, he accepted the current set of simulated questions and answers (Appendix A.13). Then, the designer asked ChatGPT to simulate interviews with the other five personas. ChatGPT was able to keep the same set of questions across all interviews. After each interview, the designer asked ChatGPT to include one more question on how the user would improve the app. The responses across the interviews showed some formulaic and mechanical response generation like the previous user research interviews. They used similar sentence structures with some key elements and terms replaced according to the specific background, routines and problems of the personas. For example, all personas "*liked the to-do list feature and reminders to take breaks and do self-care activities*" (see the underlined text for the distinct parts of responses across the interviews). However, each one of them benefited from these features in different ways aligned with their specific problems mentioned in the persona descriptions: Sarah's persona included her difficulties in managing parenting responsibilities, or Alice's persona described her feelings of being lonely and isolated.

> **ChatGPT**:
>
> **Interviewer**: Hi [Sarah/Mike/Alice], thanks for taking the time to talk to us today. Can you tell us a bit about your experience with the app associated with the first idea (time management and organization)?
>
> **Sarah**: Sure, I've been using the app for about a week now and overall I've been <u>really impressed.</u> It's been really helpful in keeping me organized and on track with my tasks and goals.
> **Interviewer**: That's great to hear. What specific features or functionality did you find most useful?

> **Sarah**: I really liked the to-do list feature and the reminders to take breaks and do self-care activities. I've been struggling with <u>balancing my work and parenting responsibilities</u>, so having a tool to help me stay organized and focused has been really valuable. …
>
> **Alice**: Sure, I've been using the app for about a week now and overall I've been <u>really happy with it</u>. It's been really helpful in keeping me organized and on track with my tasks and goals.
> **Interviewer**: That's great to hear. What specific features or functionality did you find most useful?
> **Alice**: I really liked the to-do list feature and the reminders to take breaks and do self-care activities. I've been struggling with <u>balancing my work and personal life</u>, so having a tool to help me stay organized and focused has been really valuable.

A follow-up question was asked to understand if a richer response can be obtained. Thus, the designer asked ChatGPT to ask a follow-up question to Alice. Her simulated responses were detailed and convincing.

> **Interviewer**: Can you give us some specific examples of how the self-care activities have helped you?
> **Alice**: Definitely! For example, the other day I was feeling really overwhelmed with my workload, and I used the app to recommend a yoga routine for me to try. It was just a few minutes long, but it helped me relax and clear my mind. I felt much more focused and energized afterwards, and I was able to get a lot more done. And another time, I was feeling anxious and down, and I used the app to recommend a journaling exercise. Writing down my thoughts and feelings helped me process them and feel better. It was like a mini therapy session!

The follow-up question was useful for obtaining a richer and more nuanced understanding of the user's experience. Overall, the simulated user feedback across all interviews was overwhelmingly positive. There was not much negative feedback except for some technical issues reported. After the completion of all interview simulations, the designer asked ChatGPT to generate individual interview summaries as well as an overall summary of findings across all interviews. It effectively summarized the main findings of the interviews (Appendix A.14). The following is an excerpt from the summary:

> Sarah found the app easy to use and navigate, and appreciated the app's advice and recommendations for maintaining a healthy work-life balance. She used the app's voice commands and other features regularly, and found the app's reminders to take breaks and do self-care activities helpful. However, she experienced some technical issues and found the app's interface and design could be improved. …
>
> Overall, the simulated interviews showed that the App idea 1 was well-received by users, with the majority of feedback being positive. There were some technical issues reported, but users were generally satisfied with the app's functionality and effectiveness in helping them manage their work and personal lives.

This final summary concluded the entire virtual design process with ChatGPT. The next section will discuss the capability and suitability of ChatGPT in supporting various activities, the implications on design, and future research directions.

## 4 DISCUSSION AND CONCLUSION

This study aimed to understand the capabilities, limitations and overall suitability of ChatGPT as a large language model to support the design process. Overall, ChatGPT demonstrated that it is a reasonably competent chatbot that was able to i) role-play as imaginary designers, users, and products; ii) keep the context of conversation around multiple interconnected topics over multiple turns (up to 4000 tokens); and iii) correct errors in its outputs through free form natural language conversations. However, ChatGPT also showed various limitations including difficulties in i) understanding some problems and requests; ii) keeping the context of the conversation when the session length increased; and, iii) providing rich outputs of simulated user data.

### 4.1 The Roles of ChatGPT

*ChatGPT as a designer*: ChatGPT collaborated with the first author to generate interview questions and then used the questions to simulate interviews with fictional users in a reasonably acceptable way. ChatGPT performed two interview studies: one to do user research to understand fictional user problems and needs at the beginning of the process and one to evaluate the user experience of the fictional users at the end. ChatGPT was very good at generating a list of interview questions relevant to the project topic and the stage of the design process. However, it failed to ask the full list of questions during the simulated interviews and forgot who some of the personas were. The 4000-token context history seemed a major cause. The lack of feedback on the limits of conversational context is a problem. Apart from some forgotten information on personas, ChatGPT skillfully managed conversations and understood free-form natural language prompts without requiring inputting context information each time. It successfully corrected most of its mistakes and performed follow-up requests when being asked after a particular conversational exchange. Categorically, ChatGPT can be placed in a group bringing together a complex combination of characteristics in the chatbot taxonomy: generic, open domain, generative, autonomous, and conversational [1].

*ChatGPT as a user:* ChatGPT created five fictional users and answered many questions on their behalf via some simulated interviews. The fictional users' responses were consistent with their persona descriptions, but they felt mechanically produced and repetitive. ChatGPT also answered the user experience interview questions asked to the fictional users on their behalf. Like the ones in the first interview, the responses appeared to be repetitive and mechanical, lacking the richness one would expect from actual user data. This is aligned with a recent study's findings on the lack of variability in synthetically generated user data [16]. However, in the case of ChatGPT, asking follow-up questions to obtain more detailed information was effective to break the formulaic and repetitive nature of responses as in the case of Alice's experience with the recommended self-care activities.

*ChatGPT as a product:* ChatGPT acted as a voice app when being asked to simulate some imaginary conversations that may take place between users and the app. It successfully generated such conversations by using everyday language with some vivid details aligned correctly with the desired features. The hypothetical project's choice of having a voice application relying on a conversational user interface is an important factor for ChatGPT to be able to role-play a voice application more effectively. Its capabilities for simulating non-conversational product behavior need to be investigated. Furthermore, using a conversational AI agent to envision a future product behavior can be a promising research direction. It can potentially take the role of a wizard in the conventional Wizard of Oz method where a human wizard simulates a hypothetical future system's behavior [6].

Overall, although there were some limitations, it was empowering to talk with a virtual assistant through a naturally occurring conversation (to a reasonable degree) to achieve various tasks. This form of human-AI collaboration seems to be one that will become more prevalent in the future. The two key enabling capabilities were ChatGPT's conversational competency and its notably good understanding of the problem domain. Regarding conversational competency, the fact that ChatGPT correctly understands the natural language prompts expressed in their natural form and variety was comforting and efficient. In terms of domain knowledge, LLMs' information accuracy is debatable [13]. Therefore, they may require fact-checking [17] or need to be used by people with sufficient domain knowledge for reliable use. In this hypothetical project, ChatGPT's understanding of the problem domain, its users and their needs were mostly reasonable and appropriate. However, in its current form, ChatGPT seems to be most suitable for summarizing a given content and ideating various alternative solutions. Its ability to simulate conversations can be especially useful for design and research projects on conversational applications.

There are some limitations of this study. It was performed with the research preview version of the ChatGPT on 5 December 2022. The outputs of such models get constantly re-tuned. Therefore, some ChatGPT capabilities may no longer be available or some of its limitations may be already fixed in the future. The entire interaction with ChatGPT was performed in a single lengthy session to be able to keep all the content within the scope of the conversation as much as possible. However, this may have caused ChatGPT to forget some of the important details about personas. Therefore, some of the shortcomings of ChatGPT related to forgotten information may be overcome by keeping track of session length, saving the content, and creating new sub-sessions with summarized grounding prompts.

### 4.2 Future Research and Implications

There are several areas of future research. First, future research can focus on overcoming the limitations of conversational LLMs including i) maintaining the context of conversations to be able to support longer sessions and more complex conversational inquiries [15, 30]; ii) incorporating domain-specific knowledge into the training data of LLMs to improve response relevancy and accuracy (see e.g., Codex [4]); and, iii) connecting with external resources for fact-checking [17]. Second, future research can continue with investigating, in more extensive and systematic ways, the use of LLMs to support various design activities such as user research, ideation, persona creation and interaction, prototyping and user evaluation. Third, future research can examine the ethical and social implications such as the effects of using these tools on designers' agency and capabilities; and, issues related to bias, fairness, transparency and accountability of using LLMs in the design process [12]. The studies could examine the characteristics of training data used [2, 19], and the diversity and expertise of human participants to evaluate the outputs of LLMs for reinforcement learning methods [33].

There may be several potential implications of using LLMs like ChatGPT in the design process, as it is conceivable that these technologies will be increasingly used in the future. The first author asked ChatGPT to share its opinion on potential implications and included the most sensible responses with some light editing:

*Changes in the skills and knowledge required of designers:* There may be changes in the skills and knowledge that are most valued by the design community and by employers. Designers might be expected to use these technologies with greater proficiency or to master new skills and information in disciplines like machine learning or natural language processing. Design education may also need to adjust to the changing landscape, providing courses on AI-related topics and teaching students how to integrate them into their design practice [8, 31]. Additionally, as technology advances, design teams may require more specialized skills, such as data science or programming, to improve their competitiveness.

*Changes in the design industry and the design profession:* The use of ChatGPT and other LLMs may enable designers to produce designs more quickly and efficiently, which could lead to changes in the way that design services are delivered and consumed. The increased efficiency to generate and evaluate design alternatives can lead to an increased demand for design services, and to an increased appetite for design projects that can benefit from the use of these tools. For example, clients and stakeholders may be more willing to engage designers for design projects, and to pay for design services, if they know that these services can be delivered more efficiently with fewer resources.

*An increased focus on creativity, innovation, and exploration in the design process:* Designers empowered by LLMs can explore new design possibilities and think outside the box more effectively, which could lead to an increased focus on creativity, innovation, and exploration in the design process. These tools may provide designers with a starting point for their designs, but they may also encourage designers to push the boundaries of what is possible, challenge assumptions and conventions, and explore new design approaches and paradigms.

## A APPENDICES

### A.1 Structured Personas Transcript

**Designer**: Can you create user personas without mentioning any technological solutions needed? Just describe: 1. main problem statement, 2. demographics, 3. personality traits, 4. typical daily activities, 5. main problems, 6. motivations, 7. Desires.

**ChatGPT**: Sure, here are five user personas without mentioning any technological solutions needed:

1. Main problem statement: Balancing work and parenting responsibilities while working from home
   Demographics: 32-year-old marketing manager
   Personality traits: Organized, responsible, overwhelmed
   Typical daily activities: Managing a team, attending meetings, taking care of a young child
   Main problems: Lack of time and support, difficulty staying focused and motivated
   Motivations: Providing a good example for their child, advancing their career
   Desires: A sense of accomplishment, work-life balance.

2. Main problem statement: Maintaining a healthy work-life balance while working from home
   Demographics: 45-year-old software developer
   Personality traits: Hardworking, dedicated, burnt out
   Typical daily activities: Writing code, attending meetings, working long hours
   Main problems: Overwork, lack of exercise, unhealthy habits
   Motivations: Meeting deadlines, delivering high-quality work
   Desires: A sense of accomplishment, time for themselves and their family

3. Main problem statement: Managing loneliness and isolation while working from home
   Demographics: 29-year-old graphic designer
   Personality traits: Creative, social, lonely
   Typical daily activities: Designing graphics, attending meetings, spending time alone
   Main problems: Lack of social interaction, feeling disconnected from others
   Motivations: Creating beautiful designs, being a part of a team
   Desires: Social connection, a sense of belonging

4. Main problem statement: Managing a chronic health condition while working from home
   Demographics: 55-year-old finance manager
   Personality traits: Responsible, disciplined, forgetful
   Typical daily activities: Managing a team, attending meetings, taking medication
   Main problems: Managing medication, maintaining a healthy lifestyle, tracking health goals
   Motivations: Staying healthy, providing for their family
   Desires: A healthy and active life, peace of mind

5. Main problem statement: Managing stress and improving mental health while working from home
   Demographics: 40-year-old teacher
   Personality traits: Passionate, caring, stressed
   Typical daily activities: Teaching, grading papers, attending meetings
   Main problems: Feeling overwhelmed, difficulty relaxing, poor sleep
   Motivations: Helping their students, being a good teacher

Desires: A sense of calm, and improved mental health Overall, these user personas represent a range of different needs and challenges that people working from home may face, and could be useful in designing solutions to support their health and wellbeing.

## A.2 Revised Personas Iteration 1 Transcript

**Designer**: Just a few things to add to each persona: a name, and technology use

**ChatGPT**: Sure, here are the five user personas with names and technology use added:

1. Sarah is a 32-year-old marketing manager who has been working from home for the past six months. She has a young child and struggles to balance her work and parenting responsibilities. She is an experienced technology user and is looking for a voice assistant that can help her manage her time and stay organized.
2. Mike is a 45-year-old software developer who has been working from home for over a year. He tends to work long hours and has trouble maintaining a healthy work-life balance. He is a tech-savvy user and is looking for a voice assistant that can remind him to take breaks and provide guidance on healthy habits.
3. Alice is a 29-year-old graphic designer who recently started working from home. She lives alone and finds it lonely and isolating. She is comfortable with technology and is looking for a voice assistant that can provide social support and companionship.
4. John is a 55-year-old finance manager who has been working from home for the past three years. He has a chronic health condition and struggles with managing his medication and keeping track of his health goals. He is an experienced technology user and is looking for a voice assistant that can help him manage his health and track his progress.
5. Lisa is a 40-year-old teacher who has been working from home since the start of the pandemic. She has a demanding job and often feels stressed and overwhelmed. She is familiar with technology and is looking for a voice assistant that can help her manage her stress and improve her mental health. Overall, these user personas represent a range of different needs and challenges that people working from home may face, and could be useful in designing a voice assistant to support their health and wellbeing.

## A.3 Revised Personas Iteration 2 Transcript

**Designer**: Remove any mention of voice assistant from the descriptions. just include technological devices uses regularly on a daily basis

**ChatGPT**: Sure, here are the five user personas with the mention of a voice assistant removed, and with information about the technological devices they use regularly on a daily basis:

1. Sarah is a 32-year-old marketing manager who has been working from home for the past six months. She has a young child and struggles to balance her work and parenting responsibilities. She uses a laptop, smartphone, and smartwatch on a daily daily basis.
2. Mike is a 45-year-old software developer who has been working from home for over a year. He tends to work long hours and has trouble maintaining a healthy work-life balance. He uses a desktop computer, smartphone, and smartwatch on a daily basis.
3. Alice is a 29-year-old graphic designer who recently started working from home. She lives alone and finds it lonely and isolating. She uses a laptop, smartphone, and smart speaker on a daily basis.

4. John is a 55-year-old finance manager who has been working from home for the past three years. He has a chronic health condition and struggles with managing his medication and keeping track of his health goals. He uses a laptop, smartphone, and smartwatch on a daily basis.
5. Lisa is a 40-year-old teacher who has been working from home since the start of the pandemic. She has a demanding job and often feels stressed and overwhelmed. She uses a laptop, smartphone, and tablet on a daily basis. Overall, these user personas represent a range of different needs and challenges that people working from home may face, and could be useful in designing solutions that take advantage of the technological devices they use regularly.

### A.4 Revised Personas Iteration 3 Transcript

**Designer**: Can you provide some more details about their lives in the descriptions? two more sentences in each paragraph?

**ChatGPT**: Sure, here are the five user personas with additional details about their lives:

1. Sarah is a 32-year-old marketing manager who has been working from home for the past six months. She has a young child and struggles to balance her work and parenting responsibilities. She uses a desktop computer, smartphone, and smart speaker on a daily basis. Sarah often feels overwhelmed by the demands of her job and her role as a parent, and is looking for ways to manage her time and stay organized. She enjoys spending time with her family and going for walks, and tries to make time for these activities whenever possible.
2. Mike is a 45-year-old software developer who has been working from home for over a year. He tends to work long hours and has trouble maintaining a healthy work-life balance. He uses a laptop, smartphone, and smartwatch on a daily basis. Mike is passionate about his work, but finds it hard to switch off and relax at the end of the day. He is a committed exerciser and tries to fit in a workout every day, but sometimes finds it difficult to motivate himself. He also enjoys spending time with his family and friends, and values the connections he has with them.
3. Alice is a 29-year-old graphic designer who recently started working from home. She lives alone and finds it lonely and isolating. She uses a laptop, tablet, and smart speaker on a daily basis. Alice loves her job and is passionate about creating beautiful designs, but misses the social interactions and support she used to get from her colleagues in the office. She enjoys going for walks and exploring new places, and tries to get out of the house and interact with others whenever possible. She also loves spending time with her friends and family, and values the connections she has with them.
4. John is a 55-year-old finance manager who has been working from home for the past three years. He has a chronic health condition and struggles with managing his medication and keeping track of his health goals. He uses a desktop computer, smartphone, and smartwatch on a daily basis. John is responsible and disciplined, but finds it hard to stay on top of his health goals and manage his medication. He is an avid golfer and enjoys spending time on the course, but finds it difficult to fit in exercise and healthy habits around his work schedule. He also values spending time with his family and friends, and tries to make time for these connections whenever possible.
5. Lisa is a 40-year-old teacher who has been working from home since the start of the pandemic. She has a demanding job and often feels stressed and overwhelmed. She uses a laptop, tablet, and smart speaker on a daily basis. Lisa loves teaching and is passionate about helping her students learn and grow, but finds it hard to manage her stress and maintain her mental health. She enjoys yoga and meditation, and tries to incorporate these activities into her daily routine, but sometimes finds it difficult to make time for them. She also values spending time with her family and friends, and tries to make connections and support each other whenever possible.

Overall, these user personas represent a range of different needs and challenges that people working from home may face, and could be useful in designing solutions that take into account the complex and varied lives of this population.

## A.5 Simulated Interviews with the Fictional Users Transcript

**Designer**: Ok, now simulate an interview between a designer and the first persona by asking the 10 questions you created

**ChatGPT**: Sure, here is a simulated interview between a designer and Sarah, the first persona, using the ten potential interview questions:

> **Designer**: Hi Sarah, thanks for taking the time to speak with me today. Can you tell me a bit about yourself and your work situation?
> 
> **Sarah**: Hi, sure. I'm a marketing manager and I've been working from home for the past six months. It's been a bit of a challenge balancing my work and parenting responsibilities, but I'm making it work.
> 
> **Designer**: What challenges have you faced while working from home, and how have you been coping with them?
> 
> **Sarah**: The biggest challenge for me has been trying to manage my time and stay organized. It's easy to get distracted by household tasks or my child's needs, and I find it hard to focus on my work. I've been trying to set aside specific times for work and for parenting, but it's not always easy to stick to that.
> 
> **Designer**: Are there any specific areas where you feel like you could use some support or guidance?
> 
> **Sarah**: Definitely. I would love to find a way to manage my time more effectively and stay focused on my work. I think a voice assistant could be really helpful in that regard.
> 
> **Designer**: How do you currently manage your health and wellbeing while working from home?
> 
> **Sarah**: I try to make time for exercise and for spending time with my family. I also try to eat well and get enough sleep, but it's not always easy to stick to a healthy routine.
> 
> **Designer**: What technological devices do you use regularly on a daily basis?
> 
> **Sarah**: I use a desktop computer, a smartphone, and a smart speaker on a daily basis. I find that these devices are essential for staying connected and managing my work and personal life.
> 
> **Designer**: Have you used voice assistants before, and if so, for what purposes?
> 
> **Sarah**: Yes, I've used voice assistants before for a variety of purposes. I use them to set reminders, control my smart home devices, and play music.
> 
> **Designer**: What do you like or dislike about voice assistants in general?
> 
> **Sarah**: I like that voice assistants are convenient and easy to use. I also appreciate that they can help me manage my time and stay organized. However, I do find that they can be a bit impersonal and sometimes they don't understand what I'm asking for.
> 
> **Designer**: Can you think of any specific features or functions that a voice assistant designed to support health and wellbeing of people working from home should have?
> 
> **Sarah**: I think a voice assistant designed for this purpose should be able to provide guidance on healthy habits and routines, and be able to help with time management and organization. It would also be helpful if the voice assistant could provide reminders and alerts for important tasks or appointments.
> 
> **Designer**: Are there any concerns or reservations you have about using a voice assistant in this context?
> 
> **Sarah**: I do have some concerns about privacy and security when it comes to using a voice assistant. I want to make sure that any personal information I share with the voice assistant is kept confidential and secure.

**Designer**: Is there anything else you would like to share about your experiences working from home or your needs and preferences in terms of support?

**Sarah**: I think it's important for a voice assistant designed for this purpose to be flexible and customizable. Everyone's needs and challenges are different, and a voice assistant should be able to adapt to individual users and their specific situations.

Overall, this simulated interview provides valuable insights into Sarah's experiences and needs as a person working from home, and could be useful in designing a voice assistant that meets her specific requirements.

**Designer**: Ok, now simulate an interview between a designer and the second persona by asking the 10 questions you created

**ChatGPT**: Sure, here is a simulated interview between a designer and Mike, the second persona, using the ten potential interview questions:

**Designer**: Hi Mike, thanks for taking the time to speak with me today. Can you tell me a bit about yourself and your work situation?

**Mike**: Hi, sure. I'm a software developer and I've been working from home for over a year now. It's been a bit of a challenge balancing my work and personal life, but I'm making it work.

**Designer**: What challenges have you faced while working from home, and how have you been coping with them?

**Mike**: The biggest challenge for me has been trying to maintain a healthy work-life balance. I tend to work long hours and it's hard for me to switch off and relax at the end of the day. I've been trying to set boundaries and make time for exercise and other activities, but it's not always easy to stick to that.

**Designer**: Are there any specific areas where you feel like you could use some support or guidance?

**Mike**: Definitely. I would love to find a way to manage my time more effectively and make sure I'm taking care of my health and wellbeing. I think a voice assistant could be really helpful in that regard.

**Designer**: How do you currently manage your health and wellbeing while working from home?

**Mike**: I try to make time for exercise and for spending time with my family and friends. I also try to eat well and get enough sleep, but it's not always easy to stick to a healthy routine.

Designer: What technological devices do you use regularly on a daily basis?

**Mike**: I use a laptop, a smartphone, and a smartwatch on a daily basis. I find that these devices are essential for staying connected and managing my work and personal life.

**Designer**: Have you used voice assistants before, and if so, for what purposes?

**Mike**: Yes, I've used voice assistants before for a variety of purposes. I use them to set reminders, control my smart home devices, and play music.

**Designer**: What do you like or dislike about voice assistants in general? Mike: I like that voice assistants are convenient and easy to use. I also appreciate that they can help me manage my time and stay organized. However, I do find that they can be a bit impersonal and sometimes they don't understand what I'm asking for.

**Designer**: Can you think of any specific features or functions that a voice assistant designed to support health and wellbeing of people working from home should have?

**Mike**: I think a voice assistant designed for this purpose should be able to provide guidance on healthy habits and routines, and be able to help with time management and organization. It would also be helpful if the voice assistant could provide reminders and alerts for important tasks or appointments.

**Designer**: Are there any concerns or reservations you have about using a voice assistant in this context?

**Mike**: I do have some concerns about privacy and security when it comes to using a voice assistant. I want to make sure that any personal information I share with the voice assistant is kept confidential and secure.

**Designer**: Is there anything else you would like to share about your experiences working from home or your needs and preferences in terms of support?

**Mike**: I think it's important for a voice assistant designed for this purpose to be able to adapt to the user's needs and preferences. Everyone's situation is different, and a voice assistant should be able to provide personalized support and guidance.

Overall, this simulated interview provides valuable insights into Mike's experiences and needs as a person working from home, and could be useful in designing a voice assistant that meets his specific requirements.

**Designer**: Ok, now simulate an interview between a designer and the third persona by asking the 10 questions you created

**ChatGPT**: Sure, here is a simulated interview between a designer and Alice, the third persona, using the ten potential interview questions:

**Designer**: Hi Alice, thanks for taking the time to speak with me today. Can you tell me a bit about yourself and your work situation?

**Alice**: Hi, sure. I'm a graphic designer and I recently started working from home. It's been a bit of an adjustment, but I'm enjoying the flexibility and the ability to focus on my work.

**Designer**: What challenges have you faced while working from home, and how have you been coping with them?

**Alice**: The biggest challenge for me has been the loneliness and isolation. I live alone and I miss the social interactions and support that I used to get from my colleagues in the office. I've been trying to stay connected with friends and family, and to get out of the house and interact with others whenever possible.

**Designer**: Are there any specific areas where you feel like you could use some support or guidance?

**Alice**: Definitely. I would love to find a way to stay connected and engaged with others, and to have access to support and guidance when I need it. I think a voice assistant could be really helpful in that regard.

**Designer**: How do you currently manage your health and wellbeing while working from home?

**Alice**: I try to make time for exercise and for exploring new places. I also try to eat well and get enough sleep, but it's not always easy to stick to a healthy routine.

**Designer**: What technological devices do you use regularly on a daily basis?

**Alice**: I use a laptop, a tablet, and a smart speaker on a daily basis. I find that these devices are essential for staying connected and managing my work and personal life.

**Designer**: Have you used voice assistants before, and if so, for what purposes?

**Alice**: Yes, I've used voice assistants before for a variety of purposes. I use them to set reminders, control my smart home devices, and play music.

**Designer**: What do you like or dislike about voice assistants in general?

**Alice**: I like that voice assistants are convenient and easy to use. I also appreciate that they can help me manage my time and stay organized. However, I do find that they can be a bit impersonal and sometimes they don't understand what I'm asking for.

**Designer**: Can you think of any specific features or functions that a voice assistant designed to support health and wellbeing of people working from home should have?
**Alice**: I think a voice assistant designed for this purpose should be able to provide social support and connection, and be able to help with finding activities and events to attend. It would also be helpful if the voice assistant could provide reminders and alerts for important tasks or appointments.
**Designer**: Are there any concerns or reservations you have about using a voice assistant in this context?
**Alice**: I do have some concerns about privacy and security when it comes to using a voice assistant. I want to make sure that any personal information I share with the voice assistant is kept confidential and secure.
**Designer**: Is there anything else you would like to share about your experiences working from home or your needs and preferences in terms of support?
**Alice**: I think it's important for a voice assistant designed for this purpose to be able to provide personalized support and guidance. Everyone's situation is different, and a voice assistant should be able to adapt to the user's specific needs and preferences.

Overall, this simulated interview provides valuable insights into Alice's experiences and needs as a person working from home, and could be useful in designing a voice assistant that meets her specific requirements.

**Designer**: Ok, now simulate an interview between a designer and the fourth persona by asking the 10 questions you created
**ChatGPT**: Sure, here is a simulated interview between a designer and Tom, the fourth persona, using the ten potential interview questions:
**Designer**: Hi Tom, thanks for taking the time to speak with me today. Can you tell me a bit about yourself and your work situation?
**Tom**: Hi, sure. I'm a freelance writer and I've been working from home for the past three years. It's been a bit of a challenge balancing my work and personal life, but I'm making it work.
**Designer**: What challenges have you faced while working from home, and how have you been coping with them?
**Tom**: The biggest challenge for me has been trying to maintain a healthy work-life balance. I tend to work long hours and it's hard for me to switch off and relax at the end of the day. I've been trying to set boundaries and make time for exercise and other activities, but it's not always easy to stick to that.
**Designer**: Are there any specific areas where you feel like you could use some support or guidance?
**Tom**: Definitely. I would love to find a way to manage my time more effectively and make sure I'm taking care of my health and wellbeing. I think a voice assistant could be really helpful in that regard.
**Designer**: How do you currently manage your health and wellbeing while working from home?
**Tom**: I try to make time for exercise and for spending time with my family and friends. I also try to eat well and get enough sleep, but it's not always easy to stick to a healthy routine.
**Designer**: What technological devices do you use regularly on a daily basis?
**Tom**: I use a laptop, a smartphone, and a smart speaker on a daily basis. I find that these devices are essential for staying connected and managing my work and personal life.
**Designer**: Have you used voice assistants before, and if so, for what purposes?
**Tom**: Yes, I've used voice assistants before for a variety of purposes. I use them to set reminders, control my smart home devices, and play music.

**Designer**: What do you like or dislike about voice assistants in general?

**Tom**: I like that voice assistants are convenient and easy to use. I also appreciate that they can help me manage my time and stay organized. However, I do find that they can be a bit impersonal and sometimes they don't understand what I'm asking for.

**Designer**: Can you think of any specific features or functions that a voice assistant designed to support health and wellbeing of people working from home should have?

**Tom**: I think a voice assistant designed for this purpose should be able to provide guidance on healthy habits and routines, and be able to help with time management and organization. It would also be helpful if the voice assistant could provide reminders and alerts for important tasks or appointments.

**Designer**: Are there any concerns or reservations you have about using a voice assistant in this context?

**Tom**: I do have some concerns about privacy and security when it comes to using a voice assistant. I want to make sure that any personal information I share with the voice assistant is kept confidential and secure.

**Designer**: Is there anything else you would like to share about your experiences working from home or your needs and preferences in terms of support?

**Tom**: I think it's important for a voice assistant designed for this purpose to be able to adapt to the user's needs and preferences. Everyone's situation is different, and a voice assistant should be able to provide personalized support and guidance.

Overall, this simulated interview provides valuable insights into Tom's experiences and needs as a person working from home, and could be useful in designing a voice assistant that meets his specific requirements.

**Designer**: Ok, now simulate an interview between a designer and the fifth persona by asking the 10 questions you created

**ChatGPT**: Sure, here is a simulated interview between a designer and Emma, the fifth persona, using the ten potential interview questions:

**Designer**: Hi Emma, thanks for taking the time to speak with me today. Can you tell me a bit about yourself and your work situation?

**Emma**: Hi, sure. I'm a teacher and I recently started working from home. It's been a bit of an adjustment, but I'm enjoying the flexibility and the ability to focus on my work.

**Designer**: What challenges have you faced while working from home, and how have you been coping with them?

**Emma**: The biggest challenge for me has been the loneliness and isolation. I live alone and I miss the social interactions and support that I used to get from my colleagues in the office. I've been trying to stay connected with friends and family, and to get out of the house and interact with others whenever possible.

**Designer**: Are there any specific areas where you feel like you could use some support or guidance?

**Emma**: Definitely. I would love to find a way to stay connected and engaged with others, and to have access to support and guidance when I need it. I think a voice assistant could be really helpful in that regard.

**Designer**: How do you currently manage your health and wellbeing while working from home?

**Emma**: I try to make time for exercise and for exploring new places. I also try to eat well and get enough sleep, but it's not always easy to stick to a healthy routine.

**Designer**: What technological devices do you use regularly on a daily basis?

**Emma**: I use a laptop, a tablet, and a smart speaker on a daily basis. I find that these devices are essential for staying connected and managing my work and personal life.

**Designer**: Have you used voice assistants before, and if so, for what purposes?

**Emma**: Yes, I've used voice assistants before for a variety of purposes. I use them to set reminders, control my smart home devices, and play music.

**Designer**: What do you like or dislike about voice assistants in general?

**Emma**: I like that voice assistants are convenient and easy to use. I also appreciate that they can help me manage my time and stay organized. However, I do find that they can be a bit impersonal and sometimes they don't understand what I'm asking for.

**Designer**: Can you think of any specific features or functions that a voice assistant designed to support health and wellbeing of people working from home should have?

**Emma**: I think a voice assistant designed for this purpose should be able to provide social support and connection, and be able to help with finding activities and events to attend. It would also be helpful if the voice assistant could provide reminders and alerts for important tasks or appointments.

**Designer**: Are there any concerns or reservations you have about using a voice assistant in this context?

**Emma**: I do have some concerns about privacy and security when it comes to using a voice assistant. I want to make sure that any personal information I share with the voice assistant is kept confidential and secure.

**Designer**: Is there anything else you would like to share about your experiences working from home or your needs and preferences in terms of support?

**Emma**: I think it's important for a voice assistant designed for this purpose to be able to provide personalized support and guidance. Everyone's situation is different, and a voice assistant should be able to adapt to the user's specific needs and preferences.

Overall, this simulated interview provides valuable insights into Emma's experiences and needs as a person working from home, and could be useful in designing a voice assistant that meets her specific requirements.

### A.6 Simulated Interview Summaries Transcript

**Designer**: Ok can you now summarize the first interview with Sarah?

**ChatGPT**: Sure, here is a summary of a simulated interview between a designer and Sarah, the persona you provided: Sarah is a 32-year-old marketing manager who has been working from home for the past six months. She has a young child and struggles to balance her work and parenting responsibilities. She uses a desktop computer, smartphone, and smart speaker on a daily basis.

In the interview, Sarah shared that she often feels overwhelmed by the demands of her job and her role as a parent, and is looking for ways to manage her time and stay organized. She enjoys spending time with her family and going for walks, and tries to make time for these activities whenever possible.

Sarah mentioned that she would like a voice assistant designed for supporting health and wellbeing of people working from home to be able to provide personalized support and guidance, and help with managing her time and staying organized. She also expressed concerns about privacy and security when it comes to using a voice assistant.

Overall, Sarah's interview provides valuable insights into her experiences and needs as a person working from home with young children. A voice assistant designed to support her specific requirements and preferences could be very helpful in improving her health and wellbeing.

**Designer**: Ok can you now summarize the first interview with Mike?

**ChatGPT**: Sure, here is a summary of a simulated interview between a designer and Mike, one of the personas I created: Mike is a 38-year-old IT consultant who has been working from home for the past year. He lives alone and finds it challenging to maintain a healthy work-life balance. He uses a laptop, a smartphone, and a smartwatch on a daily basis.

In the interview, Mike shared that the biggest challenge he faces while working from home is trying to maintain a healthy work-life balance. He tends to work long hours and finds it hard to switch off and relax at the end of the day. He has been trying to set boundaries and make time for activities that he enjoys, like exercise and spending time with friends and family, but it's not always easy to stick to that.

Mike mentioned that he would like a voice assistant designed for supporting health and wellbeing of people working from home to be able to provide guidance on healthy habits and routines, and help with time management and organization. He also expressed concerns about privacy and security when it comes to using a voice assistant.

Overall, Mike's interview provides valuable insights into his experiences and needs as a person working from home. A voice assistant designed to support his specific requirements and preferences could be very helpful in improving his health and wellbeing.

**Designer**: Ok can you now summarize the first interview with Alice?

**ChatGPT**: Sure, here is a summary of a simulated interview between a designer and Alice, one of the personas I created:

Alice is a 35-year-old graphic designer who has been working from home for the past three months. She lives with her partner and finds it challenging to manage her work and personal life. She uses a laptop, a smartphone, and a smartwatch on a daily basis.

In the interview, Alice shared that the biggest challenge she faces while working from home is trying to manage her time and stay organized. She tends to get distracted easily and finds it hard to stay focused on her work. She has been trying to set boundaries and make time for activities that she enjoys, like exercise and spending time with friends and family, but it's not always easy to stick to that.

Alice mentioned that she would like a voice assistant designed for supporting health and wellbeing of people working from home to be able to provide personalized support and guidance, and help with managing her time and staying organized. She also expressed concerns about privacy and security when it comes to using a voice assistant.

Overall, Alice's interview provides valuable insights into her experiences and needs as a person working from home. A voice assistant designed to support her specific requirements and preferences could be very helpful in improving her health and wellbeing.

**Designer**: Ok can you now summarize the first interview with Tom?

**ChatGPT**: Sure, here is a summary of a simulated interview between a designer and Tom, one of the personas I created: Tom is a 25-year-old software engineer who has been working from home for the past two years. He lives alone and finds it challenging to stay healthy and active while working from home. He uses a laptop, a smartphone, and a smartwatch on a daily basis.

In the interview, Tom shared that the biggest challenge he faces while working from home is trying to stay healthy and active. He tends to sit at his desk for long hours and finds it hard to make time for exercise and other healthy habits. He has been trying to set goals and stick to a routine, but it's not always easy to do that.

Tom mentioned that he would like a voice assistant designed for supporting health and wellbeing of people working from home to be able to provide personalized support and guidance, and help with managing his time and staying organized. He also expressed concerns about privacy and security when it comes to using a voice assistant.

Overall, Tom's interview provides valuable insights into his experiences and needs as a person working from home. A voice assistant designed to support his specific requirements and preferences could be very helpful in improving his health and wellbeing.

**Designer**: Ok can you now summarize the first interview with Emma?

**ChatGPT**: Sure, here is a summary of a simulated interview between a designer and Emma, one of the personas I created:

Emma is a 32-year-old teacher who has been working from home for the past six months. She lives alone and finds it challenging to stay connected and engaged with others while working from home. She uses a laptop, a tablet, and a smart speaker on a daily basis.

In the interview, Emma shared that the biggest challenge she faces while working from home is feeling lonely and isolated. She misses the social interactions and support that she used to get from her colleagues in the office, and has been trying to stay connected with friends and family and to get out of the house whenever possible.

Emma mentioned that she would like a voice assistant designed for supporting health and wellbeing of people working from home to be able to provide social support and connection, and be able to help with finding activities and events to attend. She also expressed concerns about privacy and security when it comes to using a voice assistant.

Overall, Emma's interview provides valuable insights into her experiences and needs as a person working from home. A voice assistant designed to support her specific requirements and preferences could be very helpful in improving her health and wellbeing.

**Designer**: Ok, can you synthesize the main insights from all the interview summaries?

**ChatGPT**: Sure, here are some of the main insights from the interviews with the personas:

- Many people who work from home face challenges related to maintaining a healthy work-life balance, staying organized, staying connected and engaged with others, and taking care of their health and wellbeing.
- People who work from home would like a voice assistant designed for supporting their health and wellbeing to be able to provide personalized support and guidance, help with managing their time and staying organized, and provide social support and connection.
- People who work from home have concerns about privacy and security when it comes to using a voice assistant.

Overall, these insights suggest that there is a need for a voice assistant designed specifically to support the health and wellbeing of people working from home. Such a voice assistant could be very helpful in addressing the unique challenges and needs of this group of people.

### A.7 Additional Considerations Transcript

**Designer**: Can you think of some additional challenges and obstacles to the adoption and use of such a voice assistant?

**ChatGPT**: Sure, here are some potential additional challenges and obstacles to the adoption and use of a voice assistant designed to support the health and wellbeing of people working from home:

Concerns about privacy and security: Many people are wary of using voice assistants and other digital technologies due to concerns about their privacy and security. A voice assistant designed for supporting health and wellbeing of people working from home would need to address these concerns in order to be widely adopted.

Cost and availability: Voice assistants are typically not free, and people who work from home may not have the resources or access to purchase and use them. In order to be widely adopted, a voice assistant designed for supporting health and wellbeing of people working from home would need to be affordable and widely available.

Lack of awareness and understanding: Many people may not be aware of the potential benefits of using a voice assistant for supporting their health and wellbeing, and may not understand how to use it effectively. In order to be widely adopted, a voice assistant designed for supporting health and wellbeing of people working from home would need to be well-publicized and user-friendly.

Resistance to change: Some people may be resistant to using a new technology, especially if they are used to other ways of managing their health and wellbeing. In order to be widely adopted, a voice assistant designed for supporting health and wellbeing of people working from home would need to be easy to use and provide clear benefits over other existing solutions.

Compatibility with existing technologies: Many people who work from home already have a range of digital technologies that they use on a daily basis, such as laptops, smartphones, and smart home devices. In order to be widely adopted, a voice assistant designed for supporting health and wellbeing of people working from home would need to be compatible with these existing technologies and work seamlessly with them.

User preferences and needs: Different people have different preferences and needs when it comes to their health and wellbeing. A voice assistant designed for supporting health and wellbeing of people working from home would need to be flexible and customizable in order to meet the diverse needs of its users.

Reliability and accuracy: In order to be effective, a voice assistant designed for supporting health and wellbeing of people working from home would need to be reliable and accurate. Users would need to be able to trust the information and advice provided by the voice assistant, and it would need to be able to understand and respond to their needs and preferences accurately.

Overall, there are a number of potential challenges and obstacles to the adoption and use of a voice assistant designed to support the health and wellbeing of people working from home. In order to be successful, such a voice assistant would need to address these challenges and overcome these obstacles.

### A.8 Additional Considerations Transcript

**Designer**: Generate 3 more ideas building on the ones you just suggested

**ChatGPT**: Sure, here are three more potential design ideas that could build on the ones previously suggested:

1. A voice assistant that integrates with existing time management and organization tools, such as calendars and to-do lists, and provides personalized support and guidance for managing time and staying organized. In addition to providing reminders and alerts, the voice assistant could also be used to track and monitor the user's work-life balance, and provide recommendations and suggestions for improving their work-life balance over time. For example, the voice assistant could be used to provide personalized feedback and suggestions for adjusting the user's time management and organization strategies, or to recommend other resources, such as apps or websites, that could help with maintaining a healthy work-life balance.
2. A voice-enabled social network and support group for people who work from home, where members can connect and share tips, advice, and resources for maintaining a healthy work-life balance. In addition to facilitating communication and interaction, the voice assistant could also be used to provide personalized support and guidance based on the individual needs and preferences of each user. For example, the voice assistant could be used to recommend relevant resources, such as articles, videos, and podcasts, that could help with managing stress, staying active, and maintaining a healthy work-life balance. Additionally, the voice assistant could also be used to facilitate group discussions and activities, such as virtual support groups or online workshops, that could help with building connections and fostering a sense of community among people who work from home.
3. A voice-powered virtual coach or mentor that uses artificial intelligence to provide personalized support and guidance for maintaining a healthy work-life balance, based on the individual needs and preferences of each user. In addition to providing personalized advice and recommendations, the voice assistant could also be used to track and monitor the user's progress over time, and provide ongoing support and guidance to help them maintain their healthy work-life balance. For example, the voice assistant could be used to provide personalized feedback and suggestions for adjusting the user's time management and organization strategies, or to recommend other resources, such as apps or websites, that could help with maintaining a healthy work-life balance. Additionally, the voice assistant could also be used to provide personalized reminders and alerts, such as daily check-ins or progress updates, that could help with keeping the user on track and motivated to maintain their healthy work-life balance.

These three design ideas are just examples, and there could be many other potential solutions and approaches for helping with maintaining a healthy work-life balance for people who work from home. A research-through-design project could explore and evaluate these and other ideas in order to identify the most effective and impactful solutions.

### A.9 Potential Features and Functionalities Iteration 1 Transcript

**Designer**: Envision an app prototype for the first idea. create a list of key features and functionalities of this app with examples showing how they would work

**ChatGPT**: Sure, here is a list of potential key features and functionalities for an app prototype that could help with maintaining a healthy work-life balance for people who work from home, along with examples showing how they might work:

Time management and organization tools: The app could include features such as calendars, to-do lists, and task reminders that would help users manage their time and stay organized. For example, users could create and manage their own calendars and to-do lists within the app, and the app could provide reminders and alerts to help users stay on track and meet their goals.

Personalized support and guidance: The app could provide personalized support and guidance for maintaining a healthy work-life balance, based on the individual needs and preferences of each user. For example, the app could use artificial intelligence to provide personalized recommendations and suggestions for managing time and staying organized, as well as tracking and monitoring the user's progress over time.

Social networking and support: The app could include features that would allow users to connect and interact with other people who work from home, such as forums, chat rooms, and group discussions. For example, users could share tips, advice, and resources for maintaining a healthy work-life balance, and the app could provide tools and features for facilitating communication and interaction among members.

Virtual coaching and mentoring: The app could include a virtual coach or mentor that would provide personalized support and guidance for maintaining a healthy work-life balance, based on the individual needs and preferences of each user. For example, the virtual coach could provide personalized advice and recommendations for managing time and staying organized, as well as tracking and monitoring the user's progress over time.

Integration with other technologies: The app could be designed to integrate with existing time management and organization tools, such as calendars and to-do lists, as well as other technologies, such as smart speakers, that could be used to activate and control the app using voice commands. For example, users could use voice commands to create and manage their calendars and to-do lists, or to access and interact with other features and functionalities of the app.

These are just some examples of potential key features and functionalities for an app prototype that could help with maintaining a healthy work-life balance for people who work from home. A research-through-design project could explore and evaluate these and other ideas in order to identify the most effective and impactful solutions.

### A.10 Potential Features and Functionalities Iteration 2 Transcript

**Designer**: Do it again with keeping in mind that this will be an application using voice user interface

**ChatGPT**: Sure, here is a list of potential key features and functionalities for an app prototype that could help with maintaining a healthy work-life balance for people who work from home, using a voice user interface, along with examples showing how they might work:

Time management and organization tools: The app could include features such as calendars, to-do lists, and task reminders that would help users manage their time and stay organized. Users could use voice commands to create and manage their own calendars and to-do lists within the app, and the app could provide voice-enabled reminders and alerts to help users stay on track and meet their goals.

Personalized support and guidance: The app could provide personalized support and guidance for maintaining a healthy work-life balance, based on the individual needs and preferences of each user. The app could use voice-powered artificial intelligence to provide personalized recommendations and suggestions for managing time and staying organized, as well as tracking and monitoring the user's progress over time.

Social networking and support: The app could include features that would allow users to connect and interact with other people who work from home, such as forums, chat rooms, and group discussions. Users could use voice commands to share tips, advice, and resources for maintaining a healthy work-life balance, and the app could provide voice-enabled tools and features for facilitating communication and interaction among members.

Virtual coaching and mentoring: The app could include a virtual coach or mentor that would provide personalized support and guidance for maintaining a healthy work-life balance, based on the individual needs and preferences of each user. The virtual coach could use voice-powered artificial intelligence to provide personalized advice and recommendations for managing time and staying organized, as well as tracking and monitoring the user's progress over time.

Integration with other technologies: The app could be designed to integrate with existing time management and organization tools, such as calendars and to-do lists, as well as other technologies, such as smart speakers, that could be used to activate and control the app using voice commands. Users could use voice commands to access and interact with other features and functionalities of the app, such as setting reminders or checking their progress.

These are just some examples of potential key features and functionalities for an app prototype that could help with maintaining a healthy work-life balance for people who work from home, using a voice user interface. A research-through-design project could explore and evaluate these and other ideas in order to identify the most effective and impactful solution.

### A.11 Simulated Conversations between a User and the App Transcript

**Designer**: Simulate some conversations that could happen between users and the app for each functionalities
**ChatGPT**: Here are some examples of potential conversations that could happen between users and the app for each of the functionalities described above:

- Time management and organization tools:
    **User**: "Hey, app, can you create a new calendar event for me?"
    **App**: "Sure, what would you like to name the event and when would you like it to take place?"
    **User**: "Let's call it 'Work meeting' and schedule it for tomorrow at 10am."
    **App**: "Okay, I've added a new event called 'Work meeting' to your calendar for tomorrow at 10am. Is there anything else you would like me to do?"
    **User**: "Yes, can you also add a to-do list item for me to prepare for the meeting?"
    **App**: "Sure, I've added a new to-do list item called 'Prepare for work meeting'. Is there anything else you would like me to do?"

- Personalized support and guidance:
  **User**: "Hey, app, can you give me some suggestions for managing my time and staying organized?"
  **App**: "Sure, based on your past behavior and preferences, I would recommend setting aside specific times each day for focused work, taking regular breaks to recharge, and using tools such as calendars and to-do lists to stay on top of your tasks. Is there anything else you would like me to do?"
  **User**: "Yes, can you also track my progress and provide me with regular updates on how I'm doing?"
  **App**: "Sure, I can track your progress and provide you with regular updates on how you're doing. Is there anything else you would like me to do?"

- Social networking and support:
  **User**: "Hey, app, can you connect me with other people who work from home?"
  **App**: "Sure, I can connect you with other people who work from home and share similar interests and goals. Would you like me to connect you with a specific group or topic?"
  **User**: "Yes, I would like to join a group for parents who work from home."
  **App**: "Okay, I've connected you with the group for parents who work from home. You can start chatting with other members and sharing tips and advice. Is there anything else you would like me to do?"

- Virtual coaching and mentoring:
  **User**: "Hey, app, can you provide me with some personalized advice for managing my time and staying organized?"
  **App**: "Sure, based on your past behavior and preferences, I would recommend setting aside specific times each day for focused work, taking regular breaks to recharge, and using tools such as calendars and to-do lists to stay on top of your tasks. Is there anything else you would like me to do?"
  **User**: "Yes, can you also track my progress and provide me with regular updates on how I'm doing?"
  **App**: "Sure, I can track your progress and provide you with regular updates on how you're doing. Is there anything else you would like me to do?"

- Integration with other technologies:
  **User**: "Hey, app, can you add a new calendar event for me?"
  **App**: "Sure, what would you like to name the event and when would you like it to take place?"
  **User**: "Let's call it 'Work meeting' and schedule it for tomorrow at 10am."
  **App**: "Okay, I've added a new event called 'Work meeting' to your calendar for tomorrow at 10am. Is there anything else you would like me to do?"
  **User**: "Yes, can you also add a reminder for me to prepare for the meeting?"
  **App**: "Sure, I can add a reminder for you to prepare for the meeting. What time would you like me to set the reminder for?"
  **User**: "Can you set the reminder for 9am tomorrow?"
  **App**: "Okay, I've added a reminder for 9am tomorrow to prepare for the work meeting. Is there anything else you would like me to do?"
  **User**: "No, that's all for now. Thank you, app."
  **App**: "You're welcome, happy to help. Let me know if you need anything else."

**User**: "Yes, can you also check my progress and tell me how I'm doing?"

**App**: "Sure, I can check your progress and provide you with an update on how you're doing. Based on your past behavior and goals, it looks like you are making good progress and staying on track. Keep up the good work! Is there anything else you would like me to do?"

**User**: "No, that's all for now. Thank you, app." App: "You're welcome, happy to help. Let me know if you need anything else."

### A.12 User Experience Interview Questions Transcript

**Designer**: Now, create 10 interview questions to evaluate the user experience with the app based on the first idea

**ChatGPT**: Here are 10 possible interview questions that could be used to evaluate the user experience with the app associated with the first idea:

1. What was your overall experience with the app like?
2. Did you find the app easy to use and navigate?
3. Did you find the app's advice and recommendations helpful?
4. Did you use the app's voice commands and other features regularly?
5. Did the app help you stay organized and manage your time effectively?
6. Did the app's reminders to take breaks and do self-care activities make a difference in your daily routine?
7. Did you find the app's interface and design appealing and user-friendly?
8. Did you experience any technical issues or glitches while using the app?
9. Would you recommend the app to others?
10. What, if anything, could be improved about the app?

### A.13 User Experience Interviews with the Fictional Users Transcript

**Designer**: Now, simulate an interview with the first persona who used the app associated with the first idea for a week

**ChatGPT**:

**Interviewer**: Hi Sarah, thanks for taking the time to talk to us today. Can you tell us a bit about your experience with the app associated with the first idea (time management and organization)?

**Sarah**: Sure, I've been using the app for about a week now and overall I've been really impressed. It's been really helpful in keeping me organized and on track with my tasks and goals.

**Interviewer**: That's great to hear. What specific features or functionality did you find most useful?

**Sarah**: I really liked the to-do list feature and the reminders to take breaks and do self-care activities. I've been struggling with balancing my work and parenting responsibilities, so having a tool to help me stay organized and focused has been really valuable.

**Interviewer**: Did you use the app's voice commands or other features regularly?

**Sarah**: Yes, I used the voice commands quite a bit to add tasks to my to-do list and set reminders. It was really convenient to be able to do that hands-free, especially when I was multitasking with my child.

**Interviewer**: Did you experience any technical issues or glitches while using the app?

**Sarah**: No, I didn't experience any technical issues or glitches. The app worked smoothly and was very user-friendly.

**Interviewer**: Would you recommend the app to others?

**Sarah**: Definitely. I think the app has a lot of value for anyone who is trying to balance work and home life, especially now that so many people are working from home. I would definitely recommend it to others.
**Interviewer**: Thanks for sharing your experience with us, Sarah. Is there anything else you would like to add?
**Sarah**: No, that's all. I just wanted to say that I really appreciate the value the app has provided me and I'm grateful to have had the opportunity to use it.
**Interviewer:** Is there anything you would change or add to the app to make it even better?
**Sarah**: Hmm, let me think. I think one thing that could be improved is the interface. It's user-friendly, but I think it could be made even more visually appealing and engaging. Maybe adding some color or graphics to make it more interesting to look at.
**Interviewer**: That's a great suggestion. Anything else?
**Sarah**: Well, I think it would also be useful if the app could provide some additional guidance and support for dealing with specific challenges or obstacles that people might face while working from home. For example, providing tips on how to manage distractions or communicate effectively with colleagues or family members.
**Interviewer**: Those are both excellent ideas, Sarah. Thank you for sharing them with us. We will definitely consider them as we continue to develop and improve the app.

**Designer**: Now, simulate an interview with the second persona who used the app associated with the first idea for a week
**ChatGPT**:
**Interviewer**: Hi Mike, thanks for taking the time to talk to us today. Can you tell us a bit about your experience with the app associated with the first idea (time management and organization)?
**Mike**: Sure, I've been using the app for about a week now and overall I've been pretty satisfied. It's been helpful in keeping me focused and on track with my tasks and goals.
Interviewer: That's great to hear. What specific features or functionality did you find most useful?
**Mike**: I really liked the to-do list feature and the reminders to take breaks and do self-care activities. I've been struggling with staying motivated and focused while working from home, so having a tool to help me stay organized and on track has been really valuable.
**Interviewer**: Did you use the app's voice commands or other features regularly?
**Mike**: Yes, I used the voice commands quite a bit to add tasks to my to-do list and set reminders. It was really convenient to be able to do that hands-free, especially when I was in the middle of something else. Interviewer: Did you experience any technical issues or glitches while using the app?
**Mike**: No, I didn't experience any technical issues or glitches. The app worked smoothly and was very user-friendly. Interviewer: Would you recommend the app to others?
**Mike**: Definitely. I think the app has a lot of value for anyone who is trying to stay organized and focused while working from home. I would definitely recommend it to others.
**Interviewer**: Thanks for sharing your experience with us, Mike. Is there anything else you would like to add?
**Mike**: No, that's all. I just wanted to say that I really appreciate the value the app has provided me and I'm grateful to have had the opportunity to use it.
**Interviewer:** Is there anything you would change or add to the app to make it even better?
**Mike**: Hmm, let me think. One thing that I think could be improved is the app's guidance and support for dealing with specific challenges or obstacles that people might face while working from home. For example, providing tips on how to manage distractions or communicate effectively with colleagues or family members. I think that could be really helpful.

**Interviewer**: That's a great suggestion, Mike. Anything else?

**Mike**: Well, I think it would also be useful if the app could provide some more personalized recommendations and advice based on my individual needs and goals. For example, if I'm trying to achieve a certain goal or overcome a specific challenge, the app could provide tailored advice and support to help me succeed.

**Interviewer**: Those are both excellent ideas, Mike. Thank you for sharing them with us. We will definitely consider them as we continue to develop and improve the app.

**Designer**: Now, simulate an interview with the third persona who used the app associated with the first idea for a week

**ChatGPT**:

**Interviewer**: Hi Alice, thanks for taking the time to talk to us today. Can you tell us a bit about your experience with the app associated with the first idea (time management and organization)?

**Alice**: Sure, I've been using the app for about a week now and overall I've been really happy with it. It's been really helpful in keeping me organized and on track with my tasks and goals.

**Interviewer**: That's great to hear. What specific features or functionality did you find most useful?

**Alice**: I really liked the to-do list feature and the reminders to take breaks and do self-care activities. I've been struggling with balancing my work and personal life, so having a tool to help me stay organized and focused has been really valuable.

**Interviewer**: Did you use the app's voice commands or other features regularly?

**Alice**: Yes, I used the voice commands quite a bit to add tasks to my to-do list and set reminders. It was really convenient to be able to do that hands-free, especially when I was busy with something else.

**Interviewer**: Did you experience any technical issues or glitches while using the app?

**Alice**: No, I didn't experience any technical issues or glitches. The app worked smoothly and was very user-friendly.

**Interviewer**: Would you recommend the app to others?

**Alice**: Definitely. I think the app has a lot of value for anyone who is trying to balance work and personal life, especially now that so many people are working from home. I would definitely recommend it to others.

**Interviewer**: Thanks for sharing your experience with us, Alice. Is there anything else you would like to add?

**Alice**: No, that's all. I just wanted to say that I really appreciate the value the app has provided me and I'm grateful to have had the opportunity to use it.

**Interviewer**: Is there anything you would change or add to the app to make it even better?

**Alice**: Hmm, let me think. One thing that I think could be improved is the app's guidance and support for dealing with specific challenges or obstacles that people might face while trying to balance work and personal life. For example, providing tips on how to manage time effectively or communicate with colleagues and family members. I think that could be really helpful.

**Interviewer**: That's a great suggestion, Alice. Anything else?

**Alice**: Well, I think it would also be useful if the app could provide some more personalized recommendations and advice based on my individual needs and goals. For example, if I'm trying to achieve a certain goal or overcome a specific challenge, the app could provide tailored advice and support to help me succeed.

**Interviewer**: Those are both excellent ideas, Alice. Thank you for sharing them with us. We will definitely consider them as we continue to develop and improve the app.

**Designer**: Now, simulate an interview with the fourth persona who used the app associated with the first idea for a week

**ChatGPT**:

**Interviewer**: Hi Tom, thanks for taking the time to talk to us today. Can you tell us a bit about your experience with the app associated with the first idea (time management and organization)?

**Tom**: Sure, I've been using the app for about a week now and overall I'm really pleased with it. It's been a great tool to help me stay organized and on track with my tasks and goals.

**Interviewer**: That's great to hear. What specific features or functionality did you find most useful?

**Tom**: I really liked the to-do list feature and the reminders to take breaks and do self-care activities. As a freelancer, I often find myself working long hours and getting caught up in my work, so having a tool to help me stay focused and healthy has been really valuable.

**Interviewer**: Did you use the app's voice commands or other features regularly?

**Tom**: Yes, I used the voice commands quite a bit to add tasks to my to-do list and set reminders. It was really convenient to be able to do that hands-free, especially when I was working on something that required my full attention.

**Interviewer**: Did you experience any technical issues or glitches while using the app?

**Tom**: No, I didn't experience any technical issues or glitches. The app worked smoothly and was very user-friendly.

**Interviewer**: Would you recommend the app to others?

**Tom**: Definitely. I think the app has a lot of value for anyone who is trying to balance work and personal life, especially if they work from home or have a lot of flexibility in their schedule. I would definitely recommend it to others.

**Interviewer**: Thanks for sharing your experience with us, Tom. Is there anything else you would like to add?

**Tom**: No, that's all. I just wanted to say that I'm really happy with the app and I think it has the potential to help a lot of people manage their time and stay organized. I appreciate the value it has provided me and I'm grateful to have had the opportunity to use it.

**Interviewer**: Is there anything you would change or add to the app to make it even better?

**Tom**: Well, one thing that I think could be improved is the app's ability to integrate with other productivity tools and platforms that I use regularly. For example, if I'm using a specific task management or project planning app, it would be really helpful if the app associated with the first idea could sync with it and provide even more comprehensive support for managing my time and tasks.

**Interviewer**: That's a great suggestion, Tom. Anything else?

**Tom**: Another thing that I think would be useful is if the app could provide more personalized and tailored advice and support based on my individual goals and challenges. For example, if I'm trying to achieve a specific goal or overcome a specific obstacle, it would be helpful if the app could provide tailored advice and support to help me succeed.

**Interviewer**: Those are both excellent ideas, Tom. Thank you for sharing them with us. We will definitely consider them as we continue to develop and improve the app.

**Designer**: Now, simulate an interview with the fifth persona who used the app associated with the first idea for a week
**ChatGPT**:

> **Interviewer**: Hi Emma, thanks for taking the time to talk to us today. Can you tell us a bit about your experience with the app associated with the first idea (time management and organization)?
>
> **Emma**: Sure, I've been using the app for about a week now and overall I'm really happy with it. I work from home and sometimes it can be challenging to stay organized and focused, so having a tool like this has been really helpful.
>
> **Interviewer**: That's great to hear. What specific features or functionality did you find most useful?
>
> **Emma**: I really liked the to-do list and the reminders to take breaks and do self-care activities. As a mom, I often find myself feeling overwhelmed and pulled in different directions, so having a tool to help me stay on track and take care of myself has been really valuable.
>
> **Interviewer**: Did you use the app's voice commands or other features regularly?
>
> **Emma**: Yes, I used the voice commands quite a bit to add tasks to my to-do list and set reminders. It was really convenient to be able to do that hands-free, especially when I was in the middle of something else and didn't want to stop and type on my phone.
>
> **Interviewer**: Did you experience any technical issues or glitches while using the app?
>
> **Emma**: No, I didn't experience any technical issues or glitches. The app worked smoothly and was very user-friendly.
>
> **Interviewer**: Would you recommend the app to others?
>
> **Emma**: Definitely. I think the app has a lot of value for anyone who is trying to balance work and personal life, especially if they work from home or have a lot of flexibility in their schedule. I would definitely recommend it to others.
>
> **Interviewer**: Thanks for sharing your experience with us, Emma. Is there anything else you would like to add?
>
> **Emma**: No, that's all. I just wanted to say that I'm really happy with the app and I think it has the potential to help a lot of people manage their time and stay organized. I appreciate the value it has provided me and I'm grateful to have had the opportunity to use it.
>
> **Interviewer**: Is there anything you would change or add to the app to make it even better?
>
> **Emma**: One thing that I would like to see added to the app is a feature that allows me to track my progress over time. For example, if I'm trying to work on a specific goal or habit, it would be helpful to see how I'm doing over the course of a week or month. This could be in the form of a chart or graph that shows my progress and highlights any areas where I may need to focus more effort.
>
> **Interviewer**: That's a great suggestion, Emma. Anything else?
>
> **Emma**: Another thing that I think would be useful is if the app could provide more personalized and tailored advice and support based on my individual goals and challenges. For example, if I'm trying to achieve a specific goal or overcome a specific obstacle, it would be helpful if the app could provide tailored advice and support to help me succeed.
>
> **Interviewer**: Those are both excellent ideas, Emma. Thank you for sharing them with us. We will definitely consider them as we continue to develop and improve the app.

### A.14 Summary of User Experience Interviews Transcript

**Designer**: Summarise the simulated interviews with all 5 personas for the app idea 1

**ChatGPT**: After conducting simulated interviews with the 5 personas, the following insights were gathered about their experience with the app associated with the first idea:

Sarah found the app easy to use and navigate, and appreciated the app's advice and recommendations for maintaining a healthy work-life balance. She used the app's voice commands and other features regularly, and found the app's reminders to take breaks and do self-care activities helpful. However, she experienced some technical issues and found the app's interface and design could be improved.

Mike also found the app easy to use and appreciated its advice and recommendations. He used the app's voice commands and other features regularly, and found the app helped him stay organized and manage his time effectively. However, he also experienced some technical issues and thought the app's design could be improved.

Alice found the app's tracking and analysis of her work habits helpful, but found the app difficult to navigate and had trouble using its voice commands and other features. She did not experience any technical issues, but thought the app's interface and design could be improved.

Tom had a positive overall experience with the app, and found it easy to use and navigate. He appreciated the app's advice and recommendations, and used its voice commands and other features regularly. He did not experience any technical issues, but thought the app's design could be improved.

Emma also had a positive overall experience with the app, and found it easy to use and navigate. She appreciated the app's advice and recommendations, and used its voice commands and other features regularly. She did not experience any technical issues, but thought the app's design could be improved.

Overall, the simulated interviews showed that the app idea 1 was well-received by users, with the majority of feedback being positive. There were some technical issues reported, but users were generally satisfied with the app's functionality and effectiveness in helping them manage their work and personal lives.